\documentclass[preprint]{revtex4}
\usepackage{graphicx}
\usepackage{amsmath}

\newcommand{\be}{\begin{equation}}
\newcommand{\ee}{\end{equation}}
\newcommand{\bea}{\begin{eqnarray}}
\newcommand{\eea}{\end{eqnarray}}

\begin{document}

\date{\today}

\title{Correlation between Polyakov loops oriented in two different directions
 in SU(N) gauge theory on
a two dimensional torus}
\author{Joe Kiskis}
\email{jekiskis@ucdavis.edu}
\affiliation{Department of Physics, University of California, Davis,
 CA 95616}
\author{Rajamani Narayanan}
\email{rajamani.narayanan@fiu.edu}
\author{Dibakar Sigdel}
\email{dibakar.sigdel@fiu.edu}
\affiliation{Department of Physics, Florida International University, Miami,
FL 33199.}

\begin{abstract}
We consider SU(N) gauge theories on a two dimensional torus with
finite area, $A$. Let
$T_\mu(A)$ denote the Polyakov loop operator
in the $\mu$ direction.
Starting
from the lattice gauge theory on the torus, we derive
a formula for the continuum limit of
$\langle g_1(T_1(A)) g_2(T_2(A)) \rangle$ as a function of the area of the torus where
$g_1$ and $g_2$ are class functions. We show that there exists a class function
$\xi_0$ for SU(2) such that $\langle \xi_0(T_1(A)) \xi_0(T_2(A))\rangle > 1$ for
all finite area of the torus with the limit being unity as the area of
the
torus goes to infinity. Only the trivial representation
contributes to $\xi_0$ as $A\to\infty$
whereas all representations become equally important as $A\to 0$.

\end{abstract}

\maketitle

\section{Introduction with an overview of the main result}\label{intro}

Two dimensional non-abelian gauge theories are particularly simple to
study but reveal a wealth of physics insights. Migdal~\cite{Migdal:1975zg} studied this
theory
in the context of recursion equations since these equations become
exact in two dimensions. 
Gross and Taylor~\cite{Gross:1993hu} showed that the partition
function of two dimensional QCD is a string theory.
Gross and Witten~\cite{Gross:1980he} started from the lattice theory
with the standard Wilson action on an infinite lattice and showed
factorization
to independent plaquettes prompting a possible connection between
infinite volume gauge theories and matrix model in a certain
limit~\cite{Eguchi:1982nm}.

In this paper, we start with an SU(N) lattice gauge theory on a torus 
and rewrite the theory in terms of plaquette degrees of
freedom
and two additional toron degrees of freedom. There is a constraint
imposed
by the theory being defined on a two dimensional torus and the resulting
partition
function is
\begin{align}
Z(\beta;L_1L_2) 
= \sum_r d_r {\prod_{n_1=0}^{L_1-1}\prod_{n_2=0}^{L_2-1}}
&\int\left[dU_p(n_1,n_2)\right]
f_p\left[U_p(n_1,n_2);\beta\right]\notag\\
&\int  
dT_1 dT_2 \chi_r\left[W(L_1L_2)T_2T_1T_2^\dagger T_1^\dagger \right] ,
\label{fpartfun}
\end{align}
where
\begin{itemize}
\item $\beta$ is the dimensionless lattice coupling related to the
  continuum coupling, $g$, by $\beta=\frac{1}{g^2a^2}$ where $a$ is
  the lattice spacing.
\item The single plaquette action, $f_p$, is a coupling dependent class function of the plaquette
  variable, $U_p(n_1,n_2)$.
\item $\chi_r$ is the character in the representation labelled by $r$
  and $d_r$ is the dimension of that representation. The fundamental
representation is labelled by $f$ and the trivial representation by $0$.
\item The $L_1\times L_2$ periodic lattice has $L_1L_2$ plaquettes
  with
$U_p(n_1,n_2)$; $0\le n_1 < L_1$ and $0\le n_2 < L_2$ being the
corresponding
plaquette variables.
\item $T_1$ and $T_2$ are toron variables that arise from the presence
  of
non-contractable loops on the torus.
\item The largest Wilson loop on the torus is the ordered product
\be 
W(L_1L_2)=
\prod_{n_2=0}^{L_2-1} \left[ \prod_{n_1=L_1-1}^0 U_p(n_1,n_2) \right].
\ee 
\end{itemize}

The above results on the lattice can be used to study observables in
the continuum limit on a torus of area $A$.
Ignoring a possible overall factor that does not affect computation of observables,
the continuum partition function upon integration of all variables is
of the form
\be 
Z(A) = \sum_r e^{-\frac{1}{N}C_r^{(2)} A},
\ee  
where $A$ is the dimensionless area of the torus and $C_r^{(2)}$ is the
quadratic Casimir in the $r$ representation. This is 
the starting point in~\cite{Gross:1993hu}  for the case of a torus.

Starting from (\ref{fpartfun}), we show that the continuum limit of
the expectation value of a
Wilson loop of area $0 \le X \le A$ in representation, $r$, is given
by
\be
\frac{1}{d_r} 
\left\langle \chi_r\left(W\left(X,A\right)\right)\right\rangle=\frac{1}{Z(A)}
\sum_{r',r''} \frac{n(r,r';r'')d_{r''}}{d_rd_{r'}} e^{-\frac{1}{N}C_{r'}^{(2)}(A-X)-\frac{1}{N}C_{r''}^{(2)}X} ,
\ee  
where $n(r,r';r'')$ is the number of times the representation $r''$
appears in the tensor product, $r\otimes r'$.
This coincides with the formula derived in~\cite{Aroca:1999qg} where
the techniques used for the calculation are close to the one used in
this paper.

Polyakov loop expectation values have been considered in the 
past\cite{Grignani:1995bi,Mitreuter:1996ze,Grignani:1997yg} but
the
focus has been mainly on Polyakov loop correlators in order to see
the confinement behavior. In this paper, we consider 
correlators
of two Polyakov loops oriented in the two different directions. The result
only depends on the area of the torus and the representations of the
Polyakov loops and we find
\begin{align}
M_{r_1r_2}(A)&=
\langle \chi_{r_1}(T_1(A))
\chi_{r_2}(T_2(A))\rangle
=\langle \chi_{r_2}(T_1(A))
\chi_{r_1}(T_2(A))\rangle\notag\\
&= \frac{1}{Z(A)}
\sum_r a(r_1r_2;r)  e^{-\frac{1}{N}C_r^{(2)}A},
\end{align}
where
\be
a(r_1,r_2;r)
=d_r \int   
 dT_1 dT_2 
\chi_r \left[T_2T_1T_2^\dagger T_1^\dagger
\right]
\chi_{r_1}(T_1)\chi_{r_2}(T_2)\label{polycoeff}.
\ee 
By diagonalizing the infinite dimensional matrix, $M(A)$, at
each $A$, we can find a new set of $A$ dependent orthonormal basis of class functions,
\be
\xi_i(T(A)) = \sum_r b_i^r(A) \chi_r(T(A)),\ \ \ \ 
\int dT(A) \xi^*_i(T(A)) \xi_j(T(A)) = \delta_{ij};\ \ \ \ i=0,1,\cdots
\ee
such that
\be
\langle \xi_{i}(T_1(A))
\xi_{j}(T_2(A))\rangle = \lambda_i(A) \delta_{ij}; \ \ \ \lambda_i(A) >
\lambda_{i+1}(A)\ \ \ \forall\ \ \ i.
\ee
An explicit computation for the case of $SU(2)$ results in only one
eigenvalue, $\lambda_0(A)$, satisfying the condition $\lambda_0(A) >
1$ for all finite $A$ with $\lambda_0(\infty)\to 1$ $A\to\infty$ and
$\lambda_0(A)\to\infty$ as $A\to 0$.

\section{Gauge action in terms of plaquette variables}

Consider a $L_1\times L_2$ periodic lattice. We wish to study a gauge
invariant nonabelian gauge theory on this lattice. The $(2L_1L_2)$ SU(N)
link variables are denoted by $U^g_\mu(n_1,n_2)$ for $0\le n_1 < L_1$;
$0\le n_2 < L_2$ and $\mu=1,2$. They obey periodic boundary
conditions:
\be
U^g_1(n_1,L_2)= U^g_1(n_1,0);\ \ \ \ 
U^g_2(L_1,n_2) = U^g_2(0,n_2) ;\ \ \ \ 0\le n_1 < L_1;\ \ \ \ 0\le n_2 < L_2.
\ee

Given a representative gauge field configuration, $U_\mu(n_1,n_2)$,
all configurations on this gauge orbit are given by
\bea
U^g_1(n_1,n_2) &=& g(n_1,n_2) U_1(n_1,n_2) g(n_1+1,n_2);\cr
U^g_2(n_1,n_2) &=& g^\dagger(n_1,n_2) U_1(n_1,n_2) g(n_1,n_2+1);
\eea
where $g(n_1,n_2)$ is a periodic function defined on the lattice
sites.

\begin{figure}[ht]
\centerline{
\includegraphics[width=210mm]{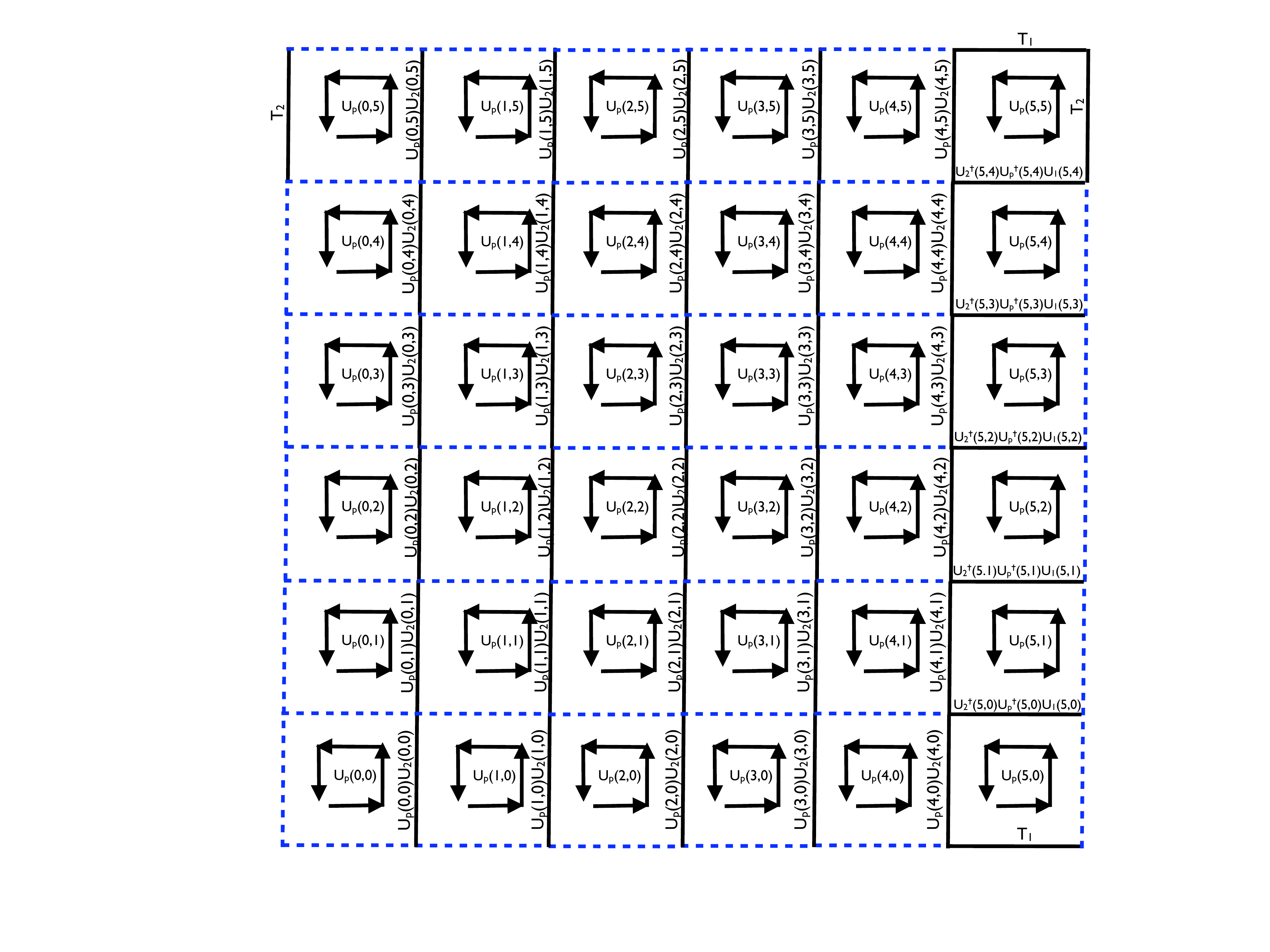}
}\caption{
A pictorial representation of the representative gauge field
configuration on a $6\times 6$ lattice. The dashed links are set to
unity.
The horizontal links are in the $1$ direction and are oriented from
left to right.
The vertical links are in the $2$ direction and are oriented from
bottom to top.}
\label{lattice}
\end{figure}

We start with the following representative gauge field configuration
(see Fig.~\ref{lattice}):
\begin{itemize}
\item $ U_1(n_1,n_2)=1;$ for $0 \le n_1 < L_1-1$ and $0 \le n_2 <
  L_2$.
\item $ U_2(0,n_2)=1;$ for $0 \le n_2 < L_2-1$.
\item $ U_1(L_1-1,0) = T_1.$
\item $ U_2(0,L_2-1) = T_2.$
\item $ U_2(n_1+1,n_2) = U_p(n_1,n_2)U_2(n_1,n_2)$ for $0 \le n_1
 < L_1-1;$ and $0 \le n_2 < L_2$.
\item $ U_1(L_1-1,n_2+1) = U_2^\dagger(L_1-1,n_2) U_p^\dagger(L_1-1,n_2)
  U_1(L_1-1,n_2)$ for $ 0\le n_2 < L_2-1$.
\end{itemize}

This configuration still has a global gauge symmetry given by $g(n_1,n_2)=g$.
The integration over all $2 (L_1L_2)$ $U_\mu^g(n_1,n_2)$ variables can
be split into 
\begin{itemize}
\item
$((L_1L_2)-1)$ $U_p(n_1,n_2)$ variables for all $0\le n_1 < L_1$;
$0\le n_2 < L_2$ except $(n_1,n_2)=(L_1-1,L_2-1)$;
\item
$((L_1L_2)-1)$ $g(n_1,n_2)$ variables for all $0\le n_1 < L_1$;
$0\le n_2 < L_2$ except $(n_1,n_2)=(0,0)$;
\item $T_1$ and $T_2$.
\end{itemize}
Note that $U_p(n_1,n_2)$ is nothing but the plaquette variable
associated with the plaquette with $(n_1,n_2)$ as the bottom-left
corner site and taken
in the counterclockwise direction (see Fig.~\ref{lattice}).  The plaquette variable,
$U_p(L_1-1,L_2-1)$, is constrained by
\be
T_1 T_2 T_1^\dagger T_2^\dagger  =
W(L_1L_2) ,
\label{t1t2c}
\ee
where
\be   
W(L_1L_2)=
\prod_{n_2=0}^{L_2-1} \left[ \prod_{n_1=L_1-1}^0 U_p(n_1,n_2) \right]
\ee   
is the largest Wilson loop on the torus 
and the product 
is path ordered.

The next step in the definition of the model is the partition function.
We assume a single plaquette action of the form
\be
e^{S_g} = \prod_{n_1=0}^{L_1-1}\prod_{n_2=0}^{L_2-1} 
f_p\left[U_p(n_1,n_2);\beta\right]
\ee
where $\beta$ is the dimensionless coupling constant on the lattice
and $f_p$ is a coupling dependent class function which 
can be expanded in terms of characters in the form
\be 
f_p\left[U;\beta\right] = \sum_r
\tilde\beta_r(\beta) \chi_r(U);\label{chwil}
\ee 
with $\tilde\beta_r(\beta)=\tilde\beta_{\bar r}(\beta)$ and real. 
The continuum limit at a fixed physical coupling, $g^2$, is obtained
by setting $\beta=\frac{1}{g^2a^2}$ and taking the lattice spacing,
$a\to 0$.
We will keep the size of the torus fixed as we take the continuum
limit by setting the dimensionless area
\be
A = \frac{L_1L_2}{\beta} = (aL_1)(aL_2)g^2\label{area}
\ee
fixed as we take $a \to 0$ and $(L_1L_2)\to\infty$.

We will use all $(L_1L_2)$ plaquette variables in our definition of the
partition function and use (\ref{t1t2c}) to restrict the integral
over $T_1$ and $T_2$.
The finite volume partition function is defined as
\be 
Z(\beta;L_1L_2) 
= {\prod_{n_1=0}^{L_1-1}\prod_{n_2=0}^{L_2-1}}\int
\left[dU_p(n_1,n_2)\right] 
f_p\left[ U_p(n_1,n_2;\beta\right]
\int   
dT_1 dT_2 \delta \left[W(L_1L_2),T_1T_2T_1^\dagger T_2^\dagger \right] .\label{partfun}
\ee 
where~\cite{drouffe}
\be
\delta(U,V) = \sum_r d_r \chi_r (UV^\dagger)\label{deltauv}
\ee
is the delta function defined on the group
with $U,V\in SU(N)$ and the sum running over all representations, $r$, with $\chi_r$
being the character and $d_r$ being the
dimension of that representation.
Insertion of (\ref{deltauv}) in (\ref{partfun}) yields the form of the
partition function, (\ref{fpartfun}), stated in Sec.~\ref{intro}.

Given that $D^r_{\alpha,\beta}(U)$ is the representation of $U$
   labelled $r$, we have the following orthogonality 
   relation~\cite{drouffe}:
\be
\int DU D^r_{\alpha\beta}(U) D^s_{\gamma\delta} (U^\dagger) =
\delta_{rs}\frac{\delta_{\alpha\delta}\delta_{\beta\gamma}}{d_r}\label{ortho}
\ee
Using (\ref{deltauv}) and (\ref{ortho}), we can
perform the integrals over $T_1$ and $T_2$ in (\ref{partfun}) to
arrive at
\be
Z(\beta;L_1L_2) 
= {\prod_{n_1=0}^{L_1-1}\prod_{n_2=0}^{L_2-1}}\int
\left[dU_p(n_1,n_2)\right]
f_p\left[U_p(n_1,n_2);\beta\right]
\left[ \sum_r \frac{1}{d_r} \chi_r(W(L_1L_2))\right]\label{plaqpfn}
\ee

Using the identity that follows from (\ref{ortho}),
\be
\int DU \chi_s(U) \chi_r(VU^\dagger W) 
= \delta_{sr}
\frac{\chi_r(VW)}{d_r},\label{chvw}
\ee
we can integrate out all $U_p(n_1,n_2)$, one after
the other, to obtain
\be
Z(\beta;L_1L_2)= 
\sum_r 
\left[
\frac{\tilde\beta_r(\beta)}{d_r}
\right]^{L_1L_2} =
\left[\tilde\beta_0(\beta)\right]^{L_1L_2}
\sum_r 
\left[
\frac{\tilde\beta_r(\beta)}{d_r\tilde\beta_0(\beta)}
\right]^{L_1L_2} 
\label{pfnform}
\ee

\section{Wilson loops}
 
Consider a $K_1K_2$ rectangular  loop with  
corners at $(0,0)$, $(K_1-1,0)$, $(0,K_2-1)$ and $(K_1-1,K_2-1)$ and 
with $0< K_1 \le L_1-1$ 
and $0< K_2 \le L_2-1$. 
As in the case of the physical size of the torus defined in (\ref{area}), 
we will keep the size of the loop fixed as we take the continuum  
limit by setting the dimensionless area of the loop
\be  
X = \frac{K_1K_2}{\beta} = (aK_1)(aK_2)g^2\label{loopsize}
\ee  
fixed as we take $a\to 0$ and $(L_1L_2)\to\infty$.

The operator is given by (see Fig.~\ref{lattice})
\be 
W(K_1K_2) = 
\prod_{i_2=0}^{K_2-1} \left[ \prod_{i_1=K_1-1}^0 U_p(i_1,i_2) \right]
\ee
 
Starting from (\ref{plaqpfn}), we have
\bea
&&Z(\beta,L_1L_2)\frac{1}{d_r} 
\langle \chi_r(W(K_1K_2))\rangle
\cr
&=&
\prod_{n_1=0}^{L_1-1}\prod_{n_2=0}^{L_2-1}\int
\left[dU_p(n_1,n_2)\right] 
f_p\left[U_p(n_1,n_2);\beta\right]
\left[ \sum_{r'} \frac{1}{d_{r'}} \chi_{r'}(W(L_1L_2))\right]\frac{1}{d_r}\chi_r(W(K_1K_2))
\cr
&&\label{wloop}
\eea 
As in the case of the partition function, we can use (\ref{chwil}) and
(\ref{chvw}) and integrate out all $U_p(n_1,n_2)$ that does not 
appear in $W(K_1K_2)$ to obtain
\bea
&& 
Z(\beta,L_1L_2)\frac{1}{d_r} 
\langle \chi_r(W(K_1K_2))\rangle
\cr
&=&
\sum_{r'} \left[
\frac{\tilde\beta_{r'}(\beta)}{d_{r'}}
\right]^{L_1L_2-K_1K_2}
\cr
&&
\prod_{n_1=0}^{K_1-1}\prod_{n_2=0}^{K_2-1}\int
\left[dU_p(n_1,n_2)\right] 
f_p\left[U_p(n_1,n_2);\beta\right]
\frac{1}{d_{r'}} \chi_{r'}(W(K_1K_2))\frac{1}{d_r}\chi_s(W(K_1K_2))
\eea 
Using the Clebsch-Gordon series~\cite{Hammermesh}, 
namely,
\be
\chi_r(U) \chi_{r'}(U) = \sum_{r''}n(r,r';r'') \chi_{r''}(U),
\ee  
where $n(r,r';'')$ is the number of times the representation, $r''$,
appears in the tensor product $r\otimes r'$,
we can perform the rest of the integrals to obtain
\be 
Z(\beta,L_1L_2)
\frac{1}{d_r} 
\langle \chi_r(W(K_1K_2))\rangle
= 
\sum_{r'} 
\left[
\frac{\tilde\beta_{r'}(\beta}{d_{r'}}
\right]^{L_1L_2-K_1K_2}
\sum_{r''} \frac{n(r,r';r'')d_{r''}}{d_rd_{r'}} \left[
\frac{\tilde\beta_{r''}(\beta)}{d_{r''}}
\right]^{K_1K_2}
\ee 
Using (\ref{pfnform}), we can write the result in the form
\be
\frac{1}{d_r} 
\langle \chi_r(W(K_1K_2))\rangle
=
\frac{ \sum_{r'} 
\left[
\frac{\tilde\beta_{r'}(\beta)}{d_{r'}\tilde\beta_0(\beta)}
\right]^{L_1L_2-K_1K_2}
\sum_{r''} \frac{n(r,r';r'')d_{r''}}{d_rd_{r'}}\left[
\frac{\tilde\beta_{r''}(\beta)}{d_{r''}\tilde\beta_0(\beta)}
\right]^{K_1K_2}
}
{\sum_{r'}
\left[
\frac{\tilde\beta_{r'}(\beta)}{d_{r'}\tilde\beta_0(\beta)}
\right]^{L_1L_2}}\label{wlform}
\ee  

One can proceed further and compute the correlations of multiple Wilson
loops where no two loops have a single plaquette in common
and show that the correlations do not depend on the separation.
This is a consequence of the form of the partition function in
(\ref{fpartfun}) where all plaquettes are independent except for a
global
constraint that only depends on the area.

\section{Polyakov loops}

In order to consider the correlation between Polyakov loops
oriented in different directions, we
start from (\ref{partfun}) and (\ref{deltauv}) and
consider expectation values of the form
\begin{align}
&Z(\beta,L_1L_2) \langle \chi_{r_1}(T_1)
\chi_{r_2}(T_2)\rangle\notag\\
=&
{\prod_{n_1=0}^{L_1-1}\prod_{n_2=0}^{L_2-1}}\int
\left[dU_p(n_1,n_2)\right]
f_p\left[U_p(n_1,n_2);\beta\right]\notag\\
&\int  
dT_1 dT_2 \left\{\sum_r
d_r\chi_r \left[W(L_1L_2)T_2T_1T_2^\dagger T_1^\dagger
\right] \right\}
\chi_{r_1}(T_1)\chi_{r_2}(T_2)
\end{align}
We can use (\ref{chwil}) and
(\ref{chvw}) and integrate out all $U_p(n_1,n_2)$ to obtain
\be
Z(\beta,L_1L_2) 
\langle \chi_{r_1}(T_1)
\chi_{r_2}(T_2)\rangle
= \sum_r a(r_1,r_2;r) \left[ \frac{ \tilde\beta_r(\beta)}{d_r}\right]^{L_1L_2} 
\ee
where
\be 
a(r_1,r_2;r)
=d_r \int    
dT_1 dT_2 
\chi_r \left[T_2T_1T_2^\dagger T_1^\dagger
\right]
\chi_{r_1}(T_1)\chi_{r_2}(T_2),\label{aseqn}
\ee  
are real coefficients.
Using (\ref{pfnform}), we can write the expectation value of the
Polyakov loops in the form
\be
M_{r_1r_2}(A)= \langle \chi_{r_1}(T_1)
\chi_{r_2}(T_2)\rangle
= \frac{
\sum_r  a(r_1,r_2;r) \left[ \frac{\tilde\beta_r(\beta)}{d_r\tilde\beta_0(\beta)}\right]^{L_1L_2} 
}
{\sum_{r}
\left[
\frac{\tilde\beta_r(\beta}{d_r\tilde\beta_0(\beta)}
\right]^{L_1L_2}}\label{plform}
\ee 
Since
\be
a(r_2,r_1;r) = a(r_1,r_2;\bar r),
\ee
it follows that $M(A)$ is a real symmetric matrix.

\section{Continuum limit}

In order to take the continuum limit, we need to take a specific
lattice action.
Since the continuum limit will not depend on the specific choice as
long as it satisfies some essential properties,
the simplest choice is the heat kernel action given by~\cite{drouffe}
\be
\tilde\beta_r(\beta) = d_r e^{-\frac{C_r^{(2)}}{N\beta}},
\ee
where $C_r^{(2)}$ is the quadratic Casimir in the $r$ representation.
In this case, $\tilde\beta_0(\beta)=1$ and
\be
\lim_{a\to 0} \left[ 
\frac{\tilde\beta_r\left(\frac{1}{g^2a^2}\right)}{d_r\tilde\beta_0\left(\frac{1}{g^2a^2}\right)}\right]
^{\frac{Y}{g^2a^2} }= 
e^{-\frac{1}{N}C_r^{(2)}Y}.\label{conlim}
\ee

\begin{itemize}
\item
The continuum limit of the partition function,
(\ref{pfnform}), is
\be
Z(A) = \sum_r e^{-\frac{1}{N} C_r^{(2)} A},
\ee 
as stated in Sec.~\ref{intro}.

\item
The continuum limit of the expectation value of the Wilson loop,
(\ref{wlform}) is
\be
\frac{1}{d_r} 
\left\langle \chi_r\left(W\left(X,A\right)\right)\right\rangle=\frac{
\sum_{r',r''} \frac{n(r,r';r'')d_{r''}}{d_rd_{r'}}
e^{-\frac{1}{N}C_{r'}^{(2)}(A-X)-\frac{1}{N}C_{r''}^{(2)}X}}
{\sum_r e^{-\frac{1}{N}C_r^{(2)} A}},
\ee 
as stated in Sec.~\ref{intro}.

\begin{itemize}
\item
Since all $C_r^{(2)} > 0$ for $r\ne 0$, it follows that
\be
\frac{1}{d_r} 
\left\langle \chi_r\left(W\left(X,\infty\right)\right)\right\rangle= 
e^{-\frac{1}{N}C^{(2)}_r X},
\ee 
which shows Casimir scaling of the string tension in the infinite area limit.
\item
Since
\be
\sum_{r''} n(r,r';r'') d_{r''} = d_r d_{r'},
\ee
it follows that
\be
\frac{1}{d_r} 
\left\langle \chi_r\left(W\left(0,A\right)\right)\right\rangle=1.
\ee 
\item
For the special case of $X=A$, we have
\be
\frac{1}{d_r} 
\left\langle \chi_r\left(W\left(A,A\right)\right)\right\rangle=
\frac{\sum_{r',r''}
\frac{n(r,r';r'')d_{r''}}{d_r d_{r'}}e^{-\frac{1}{N}C_{r''}^{(2)} A}}
{\sum_{r'} e^{-\frac{1}{N}C_{r'}^{(2)} A}}.
\ee
In the limit of $A\to\infty$, only $r'=\bar r$ contributes to the numerator
and we have
\be
\frac{1}{d_r} 
\left\langle \chi_r\left(W\left(\infty,\infty\right)\right)\right\rangle=\frac{1}{d_r^2}.
\ee
\end{itemize}
\item
The continuum limit of the correlation of Polyakov loops oriented in
two different directions, (\ref{plform}), is
\be 
M_{r_1r_2}(A)=\langle \chi_{r_1}(T_1(A))
\chi_{r_2}(T_2(A))\rangle
= 
\frac{
\sum_r a(r_1,r_2;r)  e^{-\frac{1}{N}C_r^{(2)}A}}
{\sum_{r}  
e^{-\frac{1}{N}C_r^{(2)} A}},\label{mr1r2}
\ee  
as stated in Sec.~\ref{intro}.
\begin{itemize}
\item In the limit of $A\to\infty$, only $r=0$ contributes to the sum
  in the numerator and denominator. For this special case, it follows
  from (\ref{polycoeff}) that $a(r_1,r_2;0) =
  \delta_{r_10}\delta_{r_20}$.
Therefore, 
\be
M_{r_1r_2}(\infty)
= \delta_{r_10}\delta_{r_20}.
\ee
It has one eigenvalue equal to unity and all other eigenvalues are
zero.
\item For the special case of $r_2=0$ (or $r_1=0$), we have
\be
M_{r_10}(A) = \frac{
\sum_r n(r,r_1;r)  e^{-\frac{1}{N}C_r^{(2)}A}}
{\sum_{r} 
e^{-\frac{1}{N}C_r^{(2)} A}},
\ee
\end{itemize}
\end{itemize}

\section{Eigenvalues of $M$ for the case of SU(2)}

The representations of SU(2) are labelled by $s\ge 0$ with $s$ being
an integer or an half-integer. The matrix elements obtained in
(\ref{mr1r2})
become
\be
M_{s_1s_2}(A)
= 
\frac{
\sum_s  a(s_1,s_2;s)  e^{-\frac{s(s+1)}{2}A}}
{\sum_{s} 
e^{-\frac{s(s+1)}{2} A}},\label{ms1s2}
\ee 

The selection rules for $a(s_1,s_2;s)$ defined in (\ref{aseqn}) imply
that $s_1$ and $s_2$ have to be integers. Furthermore, for a given
$s$, $a(s_1,s_2;s)$ can be non-zero only if $0 \le s_1,s_2 \le 2s$.
Therefore, if we restrict the sum in the numerator of (\ref{ms1s2}) to
$s\le S$, then we have a finite dimensional matrix of size $(2S+1)\times(2S+1)$.
The integral involved in the evaluation of $a(s_1,s_2;s)$ defined in
(\ref{aseqn}) 
can be computed using Clebsch-Gordan coefficients but we found it
easier
to perform a numerical integration by explicitly writing out $T_1$ and
$T_2$ in a fixed choice of coordinates. We can work in a gauge where
$T_1$ is diagonal. Working in the fundamental representation, we have
\begin{align}
T_1 &= \begin{pmatrix}
e^{i\eta_1} & 0 \cr
0 & e^{-i \eta_1} \cr
\end{pmatrix} & \eta_1\in [0,\pi];\notag\\
T_2 &= \begin{pmatrix}
\cos\theta_2 e^{i\alpha_2} & \sin\theta_2 e^{i\beta_2} \cr
-\sin\theta_2 e^{-i\beta_2} & \cos\theta_2 e^{-i\alpha_2}\cr
\end{pmatrix}
&\theta_2\in \left[0,\frac{\pi}{2}\right];\ \ \alpha_2,\beta_2\in [0,2\pi].
\end{align}
The eigenvalues of $T_2$ are
\be
e^{\pm i \eta_2};\ \ \ \ 
\cos\eta_2 = \cos\theta_2\cos\alpha_2;\ \ \ \ \eta_2\in [0,\pi].
\ee
The eigenvalues of $\left(T_2T_1T_2^\dagger T_1^\dagger\right)$ are
\be
e^{\pm i\eta};\ \ \ \ 
\cos\eta = 1-2\sin^2\theta_2\sin^2\eta_1;\ \ \ \ \eta\in [0,\pi].\label{etaeqn}
\ee
The explicit result for (\ref{aseqn}) is
\begin{align}
&a(s_1,s_2;s) \notag\\
=& 
\frac{2(2s+1)}{\pi} \int_0^{\pi} d\eta_1 \sin\eta_1 
\sin (2s_1+1)\eta_1
\int_0^{\frac{\pi}{2}}d\theta_2 \sin 2\theta_2 
\frac{\sin (2s+1)\eta}{\sin\eta}
\int_0^{2\pi} \frac{d\alpha_2}{2\pi}
\frac{\sin (2s_2+1)\eta_2}{\sin\eta_2}.
\end{align} 

\begin{figure}[ht]
\centerline{
\includegraphics[width=180mm]{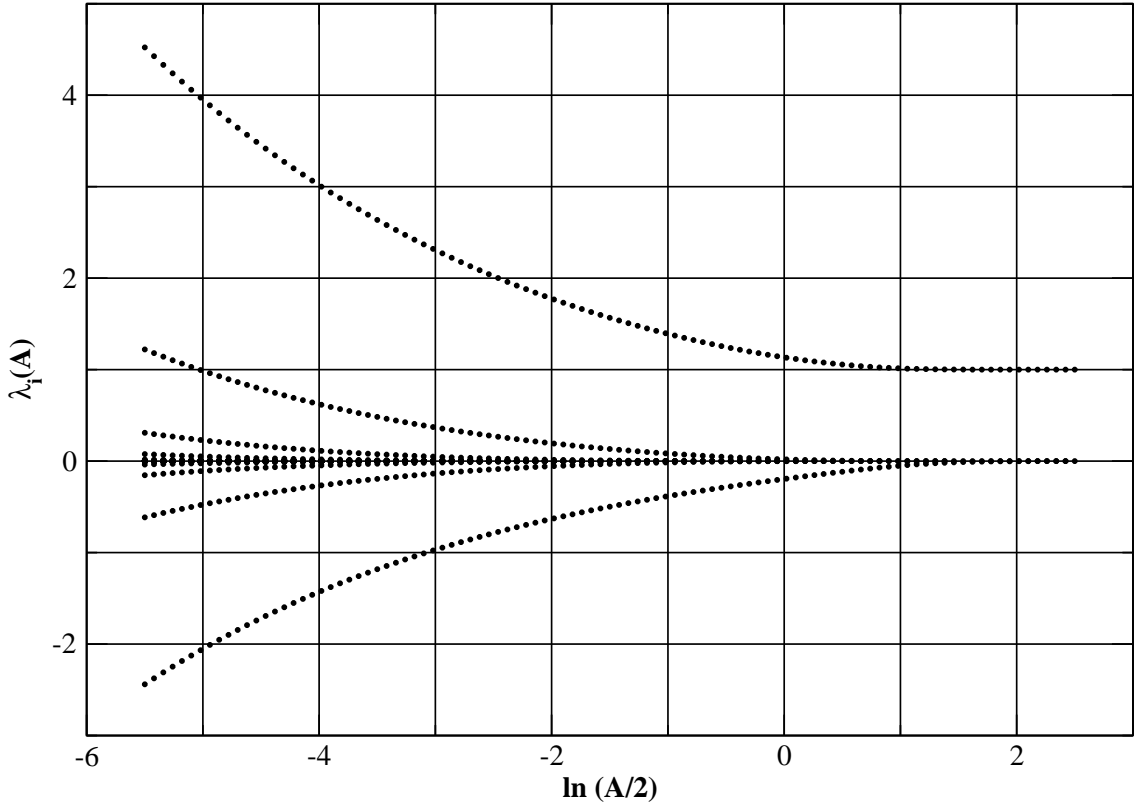}
}\caption{A plot of $\lambda_i(A)$ as a function of $A$.}
\label{eigen}
\end{figure}

Numerical results show that $|a(s_1,s_2;s)| \le 1$ and therefore it
follows that every entry in the matrix, $M(A)$, is in the range $[-1,1]$.
If we restrict the sum in the numerator of (\ref{ms1s2}) to $s\le S$,
then
we have a finite dimensional matrix which we can diagonalize and
compute all the eigenvalues. These eigenvalues will converge to
correct result and the convergence will be slower for smaller $A$.
The converged results in the range of $A\ge 10^{-3}$ are plotted 
Fig.~\ref{eigen}. The eigenvalues diverge as $A\to 0$.

\section{Discussion}

In this paper, we have studied two dimensional non-abelian gauge
theories on a torus. There is a global constraint on the plaquette
variables induced by the geometry of the torus and it only depends
on the area of the torus. We explored the area dependence on physical
observables. After showing consistency with previously known results,
we studied the correlation of two
Polyakov loops oriented in two different directions on a finite torus.
This quantity also only depends on the dimensionless area,
$A$. 
Correlations
of Polyakov loops in representations $r_1$ and $r_2$, $M_{r_1r_2}(A)$,
is a real symmetric matrix. In the large area limit,
$M_{00}(\infty)=1$ and all others are zero. This says that insertion
of
Polyakov loops in any non-trivial representation costs infinite amount
of
energy. The matrix, $M(A)$, for
SU(2)
at finite $A$ has every entry in the range $[-1,1]$. Upon
diagonalization at a fixed $A$,
we have new normalized eigenvectors of the form
\be
\xi_i(T(A)) = \sum_s b_i^s(A) \chi_s(T(A)),\ \ \   i=0,1,\cdots
\ee
with corresponding eigenvalues, $\lambda_i(A)$, satisfying
$\lambda_i(A) > \lambda_{i+1}(A)$.
Each eigenvector, $\xi_i(\theta;A)$, is an even function of $\theta\in [-\pi,\pi]$
where $e^{\pm i\theta}$ are the eigenvalues of $T(A)$ in the
fundamental
representation. The eigenvectors are normalized according to
\be
\frac{2}{\pi} \int_0^\pi d\theta \sin^2\theta \ \ \xi_i(\theta;A)
\xi_j(\theta;A) = \delta_{ij}.
\ee
Only integer valued $s$ contribute to the sum and  therefore, 
$\xi_j(\theta;A)= \xi_j(\pi-\theta;A)$.
 
The plot of the eigenvalues $\lambda_i(A)$ shown in Fig.~\ref{eigen}
has two main features:
\begin{itemize}
\item There is one eigenvalue, $\lambda_0(A) > 1$, for all finite $A$ and it
approaches unity as $A\to\infty$.
\item All other eigenvalues are less than $\lambda_0(A)$ in magnitude
  and approach zero as $A\to\infty$.
\end{itemize}
Since the expectation value of $\xi_0(T_1(A)) \xi_0(T_2(A))$ is
greater than unity, the true vacuum of the theory  contains the
insertion of
this operator. Viewed as a function of $\theta$, $\xi_0(\theta;A)$, will
develop a peak at $\theta=0$ as we decrease $A$ from infinity.

\begin{acknowledgments} 
R.N and D.S acknowledge partial support by the NSF under grant numbers 
PHY-0854744 and PHY-1205396. 
\end{acknowledgments}

\end{document}